\documentclass[onecolumn,prd,amsmath,amssymb,floatfix]{revtex4}

\usepackage{graphicx}
\usepackage{slashed}
\usepackage{amsmath,amssymb}
\usepackage{bm}
\usepackage{epsfig}

\newcommand{\bpsi}{\bar{\psi}}

\begin{document}

\title{\bf Light  vector and axial mesons effective couplings to 
constituent quarks  }

\author{ F\'abio L. Braghin }

\affiliation{$^1$ Instituto de F\'\i sica, Fed. Univ. of  Goias, P.B.131, Campus II, 
74001-970, Goi\^ania, GO, Brazil
}

\begin{abstract} 
Low energy  effective couplings of baryons' constituent quarks to light vector and axial  mesons 
are derived by considering  quark polarization for  a dressed one gluon exchange 
quark  interaction.
The quark field is splitted into two components, one for 
  background  constituent  quarks and  another one for 
quark-antiquark states, 
light mesons and the scalar  chiral condensate.
By considering a large quark effective mass derivative expansion, several 
effective coupling constants are resolved
as functions of the original model parameters and of components of 
the quark and gluon  propagators.
Besides the leading single vector meson-constituent quark gauge-type effective 
 coupling,
several
 two-vector and axial  mesons-constituent quark couplings are also obtained
in the next leading order.
Among these, 
vector and axial  mesons mixings induced by constituent quark currrents are found.
Approximated and exact ratios between the 
effective coupling constants 
 in the limit 
of very large quark effective mass  and 
 numerical estimates are exhibitted.
Numerical results of the corresponding form factors and 
of  the  (strong) 
vector   meson quadratic radius
are also presented.
\end{abstract}

\maketitle

\section{ Introduction}
\label{intro}

Light vector mesons   play an important role in a broad range of
energies in  Hadron and Nuclear Physics. 
Besides the problems related to their  structure, it is interesting to understand
in detail
the emergence of the phenomenological models describing 
their interactions with hadrons \cite{sakurai}
at different energy
density  scales
by departing  from  the more fundamental
QCD degrees of freedom.
There are 
different conceptual frameworks to describe their dynamics 
 such
as in Massive Yang Mills,  Hidden Gauge Approach
and WCCWZ
among others \cite{sakurai,HLS,birse,meissner,eLSM,anti-sym,arriola,ER-1986}
being several of them equivalent \cite{birse}.
Although  it is highly desirable  to formulate an  Effective Field Theory (EFT) 
that incorporates explicitely their degrees of freedom
some difficulties arise by trying to define the
correct  power counting rules in the framework of Chiral Perturbation theory
\cite{PLB-fuchs-etal,djukanovic-etal}.
The lightest vector mesons ( $\rho$ and $\omega$) are expected to be more relevant
for the low and intermediary energies  regimes,
being that axial chiral partners eventually are  included
 such as the  $A_1$  for the $\rho$ meson
and, less often,  an  axial partner of the $\omega$
  has also been considered 
\cite{eLSM,bloch-etal-a1b1,PDG}.
The light vector mesons effective couplings to nucleons
present less ambiguities than the 
strict (chiral)  vector mesons dynamics \cite{birse,SU3-1,tensor-current}
being extremely relevant in the short range nucleon and nuclear potentials
\cite{rho-omega,nuc-med1}.
 In the framework of the constituent quark model
 \cite{lavelle-mcm,georgi,weinberg-2010}
mesons couple directly to constituent 
quarks and the resulting coupling constants are 
proportional to the corresponding vector  mesons-baryons coupling constants.
From the criteria  of   dimensionality  and simplicity  \cite{meissner,birse}, the leading 
different   rho-quark couplings 
 can be expected to be:
$ g_v V_\mu^i (x) j_{i,v}^\mu  (x)$
and  $g_T
F_{\mu\nu}^i (x)  j_{i,T}^{\mu\nu} (x)$ 
where $F_{\mu\nu}^i (x)$ is the vector meson strength tensor.
The first one is the minimal gauge
coupling  to the vector  current   \cite{sakurai,SU3-1}
and the second one a momentum dependent 
 tensor  coupling to a tensor current
\cite{tensor-current}.
%It is also needless to emphasize that 
%vector mesons couplings to nucleons have an  important
%repulsive contribution   in the nuclear stability
%and to the  neutron-proton asymmetry  effects .
Besides that it is worth to mention vector mesons mixings are  interesting effects 
associated to charge symmetry violation \cite{thomas}
and they are   expected  to occur further  in the nuclear medium
\cite{klingl}. 
These effects are usually parameterized in terms of effective Lagrangian terms 
without  more fundamental
justification from first grounds.
So it is interesting  to search for mechanisms from 
the quark and gluon more fundamental  dynamics that 
generate  hadron and nuclear  effective models.
Instantons have been considered   to describe several
low energy  quark effective interactions
 \cite{instantons}
and other mechanisms have  also been envisaged
  \cite{EPJA-2016,PLB-2016}.
Although 
 it is possible to constraint the phenomenological couplings
usually suitable for the hadron and nuclear dynamical models, 
it is highly desirable to obtain  a  QCD-based 
 derivation  of the mechanism according to which the vector-mesons - baryons
interactions emerge. 
There are considerable  difficulties to obtain the complete
QCD effective action  by  integrating out gluons exactly from QCD
\cite{chineses}.
However
it is already interesting to understand
the emergence of hadron interactions from a restricted part of
the QCD effective action.
Besides the dynamical calculation of the hadrons
effective coupling constants it is also important to 
extract the whole momentum dependence of their interactions 
by means of form factors. 
Eventually  the  correct behavior of these effective couplings constants
and form factors 
with nuclear density,  temperature and other variables up to 
the chiral transition scales can be obtained  \cite{nuc-med2}.

In this work   light quark-antiquark vector  and axial  mesons
 couplings to (nucleons) constituent quarks are derived by considering 
a quark-quark interaction mediated by 
 a dressed (non perturbative)
gluon propagator. 
The non perturbative gluon propagator  will be therefore
an external input for the  calculation and it will be required
that it has strength enough to provide Dynamical Chiral Symmetry Breaking
(DChSB).
Besides that, this is a way of considering part of the
non Abelian gluon dynamics.
% The remaining non Abelian effects
%for example that yield intrinsic multiquark quark interactions 
%are not considered.
Therefore the quark interaction given in expression (1)
is selected   from  the QCD effective action 
to be investigated with well known analytical methods.
One loop quark polarization is calculated after  a Fierz transformation
to allow for the investigation of more complete flavor structure.
By  considering the background field  method
\cite{background} the quark field is splitted into 
sea  and background (constituent)  quarks.
Light mesons fields are introduced in the following by means of the 
the auxiliary field method \cite{ERV,PRC1,holdom,NJL}.
This procedure  has been described in detail in
 Refs. \cite{EPJA-2016,PLB-2016,FLB-2017,PRD-2014,PRD-2016}
and 
therefore in the present work it will not be described extensively.
This approach  has shown capable of providing
for example
a 
derivation of different  
constituent quark-pion effective couplings without and
with  electromagnetic couplings
and the leading terms of  chiral perturbation theory  \cite{EPJA-2016,FLB-2017}
corresponding to a whole effective field theory for low energy QCD 
%proposed by Weinberg 
\cite{weinberg-2010}.
The work is organized as it follows.
In the next Section,  the Fierz transformation, 
the quark field splitting  and the 
introduction of auxiliary meson fields  are
 briefly reminded. 
 In the following Section a 
derivative and large quark effective mass expansion 
of the sea quark determinant
is performed. 
The leading and next leading terms 
for vector mesons couplings to quarks  
are exhibited by resolving  effective coupling constants.
These effective coupling constants are
written in terms of parameters of the initial   Lagrangian 
and of components of the  quark and gluon propagators.
Some exact and approximated  ratios  between 
effective coupling constants are also exhibited for very large quark effective mass
and numerical estimates are also presented.
Finally  the   momentum dependent form factors  and a detailed investigation of 
the strong vector and axial mesons quadratic radia
are presented.
A Summary is presented in the last Section.

\section{ Flavor structure and auxiliary fields }
\label{sec:two-Q}

The dressed one gluon exchange between quarks
 can be written as \cite{ERV,PRC1}: 
\begin{eqnarray}   \label{Seff}  
Z = N \int {\cal D}[\bpsi, \psi]
\exp \;  i \int_x  \left[
\bar{\psi} \left( i \slashed{\partial} 
- m \right) \psi 
- \frac{g^2}{2}\int_y j_{\mu}^b (x) 
{\tilde{R}}^{\mu \nu}_{bc}  (x-y) j_{\nu}^{c} (y) 
+ \bpsi J + J^* \psi \right] 
\end{eqnarray}
Where the color  quark current is 
$j^{\mu}_a = \bar{\psi} \lambda_a \gamma^{\mu} \psi$, 
 $\int_x$ 
stands for 
$\int d^4 x$,
$i,j,k=0,...(N_f^2-1)$ will be used  for SU(2) flavor indices and
$a,b...=1,...(N_c^2-1)$ stands 
for color in the adjoint representation.
The sums in color, flavor and Dirac indices are implicit.
To account for the non-Abelian   structure of the gluon sector the gluon propagator
$\tilde{R}^{\mu\nu}_{bc}(x-y)$
must be non perturbative and, as an external input for the model,
it will be required to  have enough strength to yield 
Dynamical Chiral Symmetry Breaking (DChSB)
with a given strength of the quark-gluon coupling.
DChSB has been found in several works with different
approaches, though somewhat  similar,
\cite{SD-rainbow,SDE,kondo,cornwall}. 
 Other terms  from QCD effective action, such as multiquark interactions
eventually due the non Abelian gluon structure, are not 
considered.
The aim of this work is therefore to investigate
the resulting hadron effective interactions that
are obtained  by considering well known analytical  methods
presented below for the quark interaction (1).
 In several Landau type gauges 
the  gluon propagator $\tilde{R}^{\mu \nu}_{ab}(k)$
 can be written as:
 \begin{eqnarray} \label{gluonprop}
\tilde{R}^{\mu\nu}_{ab} (k) = \delta_{ab} \left[
\left( g^{\mu\nu} - \frac{k^\mu k^\nu}{ k^2}
\right) 
 R_T  (k)
+  \frac{ k^\mu k^\nu}{ k^2} 
 R_L (k) \right],
\end{eqnarray}
where $R_T(k), R_L(k)$ are
transversal and longitudinal components.
By performing a Fierz transformation  \cite{NJL}
the flavor structure 
of the interaction (\ref{Seff}) can be exploited further
by introducing the corresponding 
light quark-antiquark states and corresponding auxiliary fields
for light mesons as it is usually done for the model (\ref{Seff}) and 
NJL-type models.
This Fierz transformed interaction
will  be written in terms of the 
bilocal flavor-quark currents built with the Dirac gamma matrices and
 the flavor SU(2) Pauli matrices.
They  are given by:
$j_{s} (x,y)= \bpsi (x)\psi (y)$,
 $j_{p}  (x,y) =  \bpsi (x) i  \gamma_5 \sigma_i \psi (y)$,
$j_{si}   (x,y) =  \bpsi (x) \sigma_i  \psi (y)$,
  $j_{ps}  (x,y) =  \bpsi (x) i  \gamma_5 \psi (y)$,
$j_{V}^\mu  (x,y) =   \bpsi (x) \gamma^\mu \sigma_i \psi (y)$,
$j_{A}^\mu (x,y) =     \bpsi (x) i \gamma_5 \gamma^\mu  \sigma_i  \psi (y)$,
 $j_{vs}^\mu  (x,y) = \bpsi (x) \gamma^\mu \psi (y)$  and 
$j_{as}^\mu  (x,y) =    \bpsi (x) i \gamma_5 \gamma^\mu \psi (y)$.
The complete resulting 
set of  color singlet non local  interactions 
 are the following:
\begin{eqnarray} \label{fierz4}
\frac{ \Omega}{\alpha g^2} &\equiv& 
\left[ j_s (x,y)  j_s(y,x) + j_p^i(x,y)  j_p^i(y,x) 
+ 
j_s^i  (x,y)  j_s^i (y,x) 
\right.
\nonumber
\\
&+&  \left.
j_p(x,y)  j_p(y,x) 
 \right] R  (x-y)
-    \frac{1}{2} \left[
 j_{\mu}^i (x,y) j_{\nu}^i (y,x) 
+  {j_{\mu}^i}_A (x,y)  {j_{\nu}^i}_A (y,x)
\right.
\nonumber
\\
&+& \left.
 j_{\mu} (x,y) j_{\nu} (y,x) 
+  j_{\mu}^A (x,y)  j_{\nu}^A (y,x)
\right]  \bar{R}^{\mu\nu} (x-y) 
%\right\}
,
\end{eqnarray}
where $\alpha = 4/9$ for flavor SU(2),
 and the following kernels were defined:
\begin{eqnarray}
R {(x-y)}  &=&
3 R_T (x-y) + R_L (x-y),
\\
\bar{R}^{\mu\nu} {(x-y)}  &=&  g^{\mu\nu}
 (R_T (x-y)+R_L (x-y) )+ 2\frac{\partial^{\mu} \partial^{\nu}}{\partial^2} 
(R_T (x-y) - R_L (x-y) ).
\nonumber
\end{eqnarray}

 The background field method (BFM)
\cite{background,SWbook}
 is applied next
by splitting the  quark field   into  
sea  quark, $\psi_2$, composing light quark-antiquark states
and therefore light mesons 
and the chiral condensate,
 and the background (constituent) quark, 
$\psi_1$, eventually forming baryons.
At the one loop level 
it is enough to perform   quark bilinears shift  \cite{background}
for each of the
channel $m=s,p,si,pi,ps,V,A, as,vs$ defined with the currents above:
\begin{eqnarray}
j^m = \bpsi \Gamma^m \psi \to (\bpsi \Gamma^m \psi)_2 + (\bpsi \Gamma^m \psi)_1.
\end{eqnarray}
This separation 
   preserves chiral symmetry,
 and it may  not correspond to  a simply mode separation  of low and high energies 
which might be a too restrictive assumption. 
The ambiguity involved in this splitting 
was discussed with more details in Ref. \cite{EPJA-2016}
and it is outside the scope of this work.
The effective Fierz transformed interaction $\Omega$
is then  rewritten as sum of the different quark components interactions
as $\Omega = \Omega_1 + \Omega_2 + \Omega_{12}$,
where $\Omega_i$ stands for each of the component and $\Omega_{12}$ 
for the interaction terms between $\psi_1$ and $\psi_2$.
Instead to proceed by neglecting $\Omega_2$ according to
 the usual one loop BFM,
the auxiliary field method 
is  considered
to make possible the functional integration of the 
sea quark field.
Besides that, it allows for introducing light mesons fields.
A set of bilocal  auxiliary fields (a.f.)  is introduced by means of unitary functional integrals
multiplying the generating functional
 \cite{kleinert}.
Although this work is concerned only with the vector and axial  mesons,
all the a.f. will be introduced and, latter, some of them will be neglected.
There is one  bilocal a.f. associated to each of the quark currents, and they
are  the following:
$S(x,y), P_i(x,y), S_i(x,y), P(x,y), V_\mu^i(x,y), V_\mu(x,y), \bar{A}_\mu^i(x,y)$
and $\bar{A}_\mu(x,y)$.
This way, besides the 
rho and A$_1$ mesons, the isoscalar vector
$\omega$ and an  isoscalar axial $f_1$ \cite{PDG,eLSM} are also considered,
besides  a scalar iso-quartet ($S_i$ and $P$)
that will be  taken into account elsewhere.
The (unity Jacobian) shifts in the functional  integrals   
also  generate couplings to sea   quarks.
The bilocal auxiliary fields   give origin to punctual  meson fields
by expanding in an infinite basis of local meson  fields \cite{PRC1}, for instance 
a particular bilocal field the  vector $V_\mu^i (x,y)$
can be writtten in terms of a corresponding
 complete orhonormal
sum of local fields  as:
\begin{eqnarray} \label{vec-mes}
V^\mu_i (x,y) = V^\mu_i  \left( \frac{x+y}{2}, x-y \right)
= V^\mu_i (u ,z) =  
 \sum_k F_{k} (z) V^\mu_{i,k} (u),
\end{eqnarray}
where $ F_{k}$ are vacuum functions invariant under translation for
 each of the   local field $V^\mu_{i,k} (u)$.
For the low energy regime one  might
 pick up only the lowest energy modes, lighest $k=0$
and making the form factors  
to reduce to constants in the zero momentum  limit
$F_{k}(z) = F_{k} (0)$.
In the
 case of expression (\ref{vec-mes})
this mode turns out to be structureless isotriplet 
local mesons
$V^\mu_{i,k=0} (u) = \rho^\mu_i (x)$.
From here on, these structureless
 lowest modes for each of the channels will be denoted
by:
$S(x), P_i(x), S_i(x), P(x), V_\mu^i(x), V_\mu(x), \bar{A}_\mu^i(x),
\bar{A}_\mu (x)$.
In the present work only 
the vector and axial  mesons couplings to  constituent quarks will be addressed.
Pions and the other eventual scalar or pseudoscalar quark-antiquark states
have been considered elsewhere and will be neglected, 
except for the fact that the scalar field can give rise to a constant
contribution in the vacuum.
The resulting structureless vector and axial mesons local couplings to 
 quarks, by omitting the index $_2$, can be written as:
\begin{eqnarray} \label{vectors-q}
\bpsi (x) \Xi_v (x-y) \psi  (y) = 
\bpsi (x) \tilde\Xi_v (x)  \delta(x-y) \psi  (y)  &=& 
 - \bpsi (x)
 \frac{\gamma^\mu }{2}
 \left[ 
 F_v  \sigma_i
 \left(     V_{\mu}^i (x)
+  i  \gamma_5 
 \bar{A}_{\mu}^i (x) \right)
\right.
\nonumber
\\
&+&
\left.  F_{vs}   
 \left(     V_{\mu} (x)
+  i  \gamma_5 
 \bar{A}_{\mu} (x) \right) 
  \right] \psi (x)  \delta (x-y)
,
\end{eqnarray}
where 
the  constants $F_v$ and 
$F_{vs}$ provide   the canonical field definition
respectively  of
rho and A$_1$ mesons and 
of
$\omega$ and axial  $f_1$.

The  Gaussian integration of the 
sea  quark field can now be performed and,
by making use of the identity
$\det A = \exp \; Tr \; \ln (A)$, 
it yields:
\begin{eqnarray} \label{Seff-det}  
S_{eff}   &=&   Tr  \; \ln \; \left\{
- i \left[ {S_0}^{-1} (x-y) 
+ \Xi_v  (x-y)
+
\sum_q  a_q \Gamma_q j_q (x,y) \right]
 \right\} 
\end{eqnarray}
where 
$Tr$ stands for traces of all discrete internal indices 
and integration of  spacetime coordinates, 
$Tr = \int d^4x \, tr_D \, tr_C \, tr_F$
with the traces in Dirac, color and flavor indices,
 where the free quark kernel can be written as 
$S_0^{-1} (x-y) = \left(  i \slashed{\partial} -  m
\right) \delta (x-y)$ 
and the following notation was used for the 
constituent quark currents,
by omitting  
 the index $_1$
since sea  quarks have already been integrated out:
\begin{eqnarray} \label{Rq-j}
\frac{\sum_q  a_q \Gamma_q  j_q (x,y)}{ \alpha g^2}
&=&
2   R (x-y)
 \left[  \bpsi (y) \psi(x)
+ i  \gamma_5 \sigma_i  \bpsi (y) i \gamma_5  \sigma_i \psi (x)
\right. 
\nonumber
\\
&+& \left.   \bpsi (y) \sigma_i \psi(x)
+ i  \gamma_5  \bpsi (y) i \gamma_5  \psi (x)
\right]
\nonumber
\\
&-& 
 \bar{R}^{\mu\nu} (x-y) \gamma_\mu  \sigma_i \left[
 \bpsi (y) \gamma_\nu  \sigma_i \psi(x)
+  i \gamma_5   \bpsi  (y)
i \gamma_5 \gamma_\nu  \sigma_i \psi (x) \right]
\nonumber
\\
&-& 
 \bar{R}^{\mu\nu} (x-y) \gamma_\mu   \left[
 \bpsi (y) \gamma_\nu  \psi(x)
+  i \gamma_5   \bpsi  (y)
i \gamma_5 \gamma_\nu   \psi (x) \right] .
\end{eqnarray}
Expression (\ref{Seff-det}) has already  been investigated 
in different limits:
for the complete pion sector by considering
pion structure for example in \cite{PRC1} and structureless pions coupled to constituent quarks
in the vacuum and coupled to the electromagnetic field
 \cite{EPJA-2016,FLB-2017} 
providing all the leading
chirally symmetric and symmetry breaking
terms of Chiral Perturbation Theory. 
In Refs. \cite{meissner,ER-1986}
 the chiral vector mesons sector
($\rho$ and $A_1$), without quarks, was investigated at length.
The purely constituent quark sector was investigated 
to derive higher order quark effective couplings,
and magnetic field dependent effective interactions 
\cite{PLB-2016,PRD-2016,FLB-2017}.

The determination of the auxilary fields in the ground state makes possible to 
incorporate dynamical chiral symmetry breaking,
as it is usually done for the NJL model and analogously for  the 
Schwinger-Dyson approach.   
 The  saddle point equations for the a.f.,   
 by denoting each of them  by $\phi_\alpha$, are given by:
$\frac{\partial S_{eff}}{\partial \phi_\alpha} = 0$.
These equations for the NJL model and global color model
  have been analyzed in many works
in the vacuum or under a finite energy density.
The scalar a.f. is the only  non trivially zero in the vacuum
provided the strength of the gluon propagator  
and the quark gluon (running) coupling constant
are  strong enough for that.
It corresponds to a scalar quark-antiquark
condensate and it produces a large effective quark mass.
A constant value for the solution of the scalar field gap equation
  yields  a correction to the 
 quark effective mass  in expression (\ref{Seff-det}).
In this case
the  quark kernel above can  then  be   written in terms of the quark effective mass 
$M^*$ as usually done \cite{EPJA-2016,PLB-2016,FLB-2017}
and besides that it might  incorporate  the  quark coupling to 
vector and axial mesons that might be 
seen as  a covariant derivative, $\slashed{D}_v = \slashed{\partial} -  i 
\; \tilde\Xi_v$.
It 
can then be written as:
\begin{eqnarray} \label{S0v}
S_{v}^{-1}(x-y)  =  
\left(
  i \slashed{\partial}
-
  M^*  \right)
\delta (x-y)
+ \Xi_v (x-y) 
.
\end{eqnarray}

\section{ Large quark mass expansion and effective couplings}

  The quark  determinant can be rewritten as:
\begin{eqnarray} \label{Idet}
I_{det} &=&  
\frac{i}{2} Tr \ln 
\left[  \left( 1  
+
S_{v} (x-y)
\sum_q a_q \Gamma_q j_q (y,x) \right)
\right.
\nonumber
\\
&& \left.
 \left( 1  +  
{S}_{v} (x-y) 
 \left(\sum_q \bar{a}_q \Gamma_q j_q  (y,x) \right)
 \right)^*
\right]
+ I_0  ,
\end{eqnarray}
where 
$I_0$  yields a multiplicative constant in the generating functional
with corrections exclusively from the vector and axial mesons
 \cite{meissner}  which 
are outside the scope of this work.

Next, a 
 large quark and gluon effective masses
 and zero order derivative  expansion of  the determinant is performed 
at the zero order derivative expansion \cite{mosel}. 
Equivalently a weak vector/axial field and large gluon effective mass
can be considered.
The large gluon effective mass limit leads to a weak strength of the 
gluon propagator in expression (\ref{Idet}) and consequently 
small effective coupling constants  for the terms of the expansion with quark currents 
as shown below.
The leading effective action interaction terms $I_{det}^{l.o.}$
for the  effective constituent quark couplings to the 
canonically defined vector and axial mesons are presented now.
As an example the vector meson $V_\mu^i$ interaction with 
constituent quark term is given by  the following effective action term:
\begin{eqnarray}
I_{det, V_\mu^i}^{l.o.} =  \frac{i}{2} \; Tr \;
\left(  S_0 (y-x) \frac{\gamma^\mu \sigma_i}{2} V_\mu^i (x)
S_0 (x-z) \bar{R}^{\rho\sigma} (y-z) \gamma_\rho \sigma_j 
\bpsi(z) \gamma_\sigma \sigma_j \psi(y) \right) .
\end{eqnarray}
With the insertion of complete sets of orthogonal momentum
states, 
 effective coupling constants, 
$g_{r1}$ and $g_{v1}$, are resolved in 
the local  limit for  zero momentum exchange
and the term above can be written as 
$I_{det, V_\mu^i}^{l.o.} = 
\int d^4 x \; g_{r1} V^\mu_i
\bpsi \gamma_\mu \sigma_i \psi$. 
In this limit 
 the resulting four  leading  effective Lagrangian interaction terms
can be written as:
\begin{eqnarray}
\label{vec-quark}
{\cal L}_{v-q}
=
g_{r1}
\left( 
V^{\mu}_i  (x) j^{V,i}_\mu (x)
+ 
\bar{A}^\mu_i (x) j^{A,i}_\mu (x)
\right)
+ g_{v1} 
\left( 
V^\mu  (x) j_\mu  (x)
+ 
\bar{A}_\mu (x)  j_A^\mu (x)
\right)
\end{eqnarray}
where, by taking the traces in Dirac, color and isospin indices,
the following effective dimensionless coupling constants
 were defined
\begin{eqnarray} \label{gr1}
g_{r1} = g_{v1}  &=& 4 i   
N_c  d_1   (\alpha g^2)
\; Tr' \; 
(( {S}_0  (k) S_0 (k)
\bar{\bar{R}} (k) ))
,
\end{eqnarray}
where 
$\bar{\bar{R}} (k)  = \bar{R}^{\mu\nu} (k) \; g_{\mu\nu}$,
 $Tr'$ stands for the integral in internal momentum
of components  of quark and gluon kernels for the limit of 
zero momentum exchange
and 
$d_n =   i \frac{(-1)^{n+1} }{2 n}$.
These couplings correspond to the minimal couplings proposed by
Sakurai  \cite{sakurai} and  extended for the chiral case with axial mesons.
Differently from the nucleon level, the 
$\rho$-constituent quark and $\omega$-constituent quark
coupling constants are equal without a factor $1/3$ for the rho coupling
\cite{rho-omega}.
The corresponding Feynman diagrams are  presented in Figure (1a) 
where the dashed line stands for any of the vector or axial 
mesons.
The  zero order derivative expansion is suitable for 
 the local  low energy regime and  
higher  order derivative effective couplings 
of the 
type:
$G_{dnVj} \partial^{n} V_\mu (x)  j^\mu  (x)$ for 
$n  \geq 2$ are neglected.

Besides that, in the next 
leading order of the large quark mass expansion,
 there are three types of 
 two-vector/axial -mesons constituent quark current couplings.
The very longwavelength and zero momentum transfer limits, for the 
 leading  terms from the first order expansion, 
 can be written, in terms of the canonically normalized mesons and by 
omitting the spacetime dependence of the fields and currents, as:
\begin{eqnarray} \label{2v-j}    
{\cal L}_{2v-j} &=&  
 g_{va-j}
\left( 
\left[  V_\mu   \bar{A}^\mu_i 
+  V_\mu^i   \bar{A}^\mu
\right]
 j_{ps}^i 
+
\left[   V_\mu^i  \bar{A}^\mu_i  
+
V_\mu  \bar{A}^\mu  
\right]
j_{ps}   \right)
\nonumber
\\
&+& g_{va-j} \left( 
\left[
 ( V_\mu^2  + \bar{A}_\mu^2 ) 
 + 
 ( {V_{\mu}^i}^2   +  (\bar{A}^\mu_i    )^2 ) 
\right] j_s  
+ 
\left[V^\mu  V_{\mu}^i   + \bar{A}_\mu  \bar{A}^\mu_i \right]
j_s^i 
\right)
\nonumber
\\
&+&
g_{F js} 
\left[
\left(
( {\cal F}_{\mu\nu}  {\cal F}^{\mu\nu} 
+ 
{\cal F}_{\mu\nu}^A   {\cal F}^{\mu\nu}_A 
)
+
( {\cal F}_{\mu\nu}^i  {\cal F}^{\mu\nu}_i 
+ 
{\cal F}_{\mu\nu}^{A,i}   {\cal F}^{\mu\nu}_{A,i} 
)
 \right)
j_s 
\right.
\nonumber
\\
&+& 
  \left.
\left( {\cal F}_{\mu\nu}^i   {\cal F}^{\mu\nu}_{A} 
+ {\cal F}_{\mu\nu}  {\cal F}^{\mu\nu}_{A,i} 
\right)
j_p^i 
+
\left(  {\cal F}_{\mu\nu}^i  {\cal F}^{\mu\nu} 
+
{\cal F}_{\mu\nu}^{A,i}  {\cal F}^{\mu\nu}_{A}
\right)
j_s^i 
+
\left(
 {\cal F}_{\mu\nu}  {\cal F}^{\mu\nu}_{A} 
+ {\cal F}_{\mu\nu} {\cal F}^{\mu\nu}_A \right)
j_p 
\right]
\nonumber
\\
&+& 
g_{\epsilon v}
i \epsilon_{ijk}
\left(
{\cal F}_{\mu\nu}^i  V^\nu_j 
+ {\cal F}^{i,A}_{\mu\nu}  \bar{A}_j^\nu 
\right)
 j_{k,V}^\mu 
+
g_{\epsilon v}
i \epsilon_{ijk}
\left(
{\cal F}_{\mu\nu}^i   \bar{A}^\nu_j 
+ {\cal F}^{i,A}_{\mu\nu}   \bar{V}_j^\nu 
\right)
j_{k,A}^\mu 
\end{eqnarray}
where the Abelian tensors for each of the fields were defined as:
\begin{eqnarray} \label{tensors}
{\cal F}^i_{\mu\nu}
&=& 
 \partial_\mu V_\nu^i - \partial_\nu V_\mu^i
,
\;\;\;\;\;\;\; {\cal F}_{\mu\nu} =  \partial_\mu V_\nu - \partial_\nu V_\mu,
\\ 
{\cal G}^i_{\mu\nu}
&=& 
 \partial_\mu \bar{A}_\nu^i - \partial_\nu \bar{A}_\mu^i
,
\nonumber
\;\;\;\;\;\;\;
{\cal G}_{\mu\nu}  
=  \partial_\mu \bar{A}_\nu - \partial_\nu \bar{A}_\mu .
\end{eqnarray}
 In the expression  (\ref{2v-j})
the  following effective coupling constants have been defined
for the zero momentum transfer limit:  
\begin{eqnarray} \label{gva1}
 g_{va-j} 
% = -  M^* g_{FV}
 &= &  
i \; 
12 N_c  d_2   
(\alpha g^2)
\;
Tr' (( {S}_0 (k) {S}_0 (k)  {S}_0 (k)  R (k) ))
, 
\\   \label{FV}
g_{\epsilon v} &=&
- \frac{ g_{va-j}}{M^*}  ,
\\   \label{Fjs}
g_{Fjs} &=&
-   i \;   12 N_c d_2     (\alpha g^2)\; Tr'
 (( 
 {S}_0 (k)
\tilde{S}_0 (k)
\tilde{S}_0 (k)
R(k)  
 \;
))
,
\end{eqnarray}
In  these  expressions 
the following function is  used:
$\tilde{S}_0 (k)  = \frac{1}{ k^2 - {M^*}^2 }
$, by implicitely assuming a regularization
procedure.
Interesting to note there are very few coupling constants for 
several different couplings along   the lines of the Universality idea.
Whereas the single vector/axial meson coupling to quark current
is dimensionless, these couplings have the following dimensions:
$g_{va-j} \sim M^{-1}$, $g_{\epsilon v} \sim M^{-2}$ and 
$g_{Fjs} \sim M^{-3}$ where $M$ is an energy scale.
%{\it
 These effective coupling constants are
the zero momentum exchange limit of  
the corresponding form factors
 and they are proportional to the
elastic and inelastic  scattering amplitudes.
%{\bf
 The   inelastic case  terms  might also  be seen as type  of mixing
 mediated or induced by different quark 
currents. 
Among these
effective coupling constants, 
$g_{Fjs}$ is  one order of magnitude smaller 
than the others  within the large quark mass $M^*$ expansion
 by a factor 
$\tilde{S}_0 \sim 1/{M^*}^2$.
Most of these couplings do not seem to be possibly 
 incorporated
into a chiral vector/axial mesons gauge framework 
\cite{meissner,birse,HLS}
in the sense that 
the couplings of the first line ($g_{va-j}$) and also from the last line
($g_{\epsilon v}$), such as $V_\mu \bar{A}^\mu_i j_{ps}^i$,
cannot be written in terms of parts of non Abelian vector mesons tensors
that generalize  expressions (\ref{tensors}).
In progressively higher order terms of the determinant expansion,
i.e. for more than two vector mesons involved or more than one 
constituent quark current,
there are also  higher order n -vector mesons-quark interactions
being that each of the additional external vector/axial  meson field 
has an extra factor $S_0(k)$ or $\tilde{S}_0$(k) that 
contributes to suppress the corresponding coupling  constant 
 by $1/M^*$ or $1/{M^*}^2$ in the large quark mass expansion.
The strengths of the corresponding
higher order vector/axial  mesons   couplings to constituent quarks
are reduced considerably and progressively 
in the limit of large quark effective mass.
Exactly the same behavior was noticed for higher order 
quark effective interactions \cite{PLB-2016}.
At higher energies eventually close to the 
chiral restoration transition 
 the large quark effective mass expansion 
might  not be  valid anymore and a different treatment of the determinant is required.
%In the present work, the vector/axial meson structures 
%are associated to the vector quark currents $\bpsi \gamma_\mu \Gamma_i \psi$
%(where  $\Gamma_i$ are possibly isospin and or $\gamma_5$ operators).
Other types of quark
 currents 
might be considered to contribute to the structure of a  vector meson
 such as:
$ j_{V_1}^{\mu} = \bpsi i \partial^\mu \psi$ 
and 
$
j_{T}^{\mu\nu} = \bpsi \sigma^{\mu\nu} \psi$.
Although the first of them might be obtained from the 
derivative expansion, their contributions for the 
vector mesons structure and their interactions
with  constituent quark interactions
are outside the scope of this work.

The corresponding Feynman diagrams are presented in 
figure (1b)
 for the effective couplings with $g_{va-j}$;
(1c) for the effective couplings 
of the types of $g_{\epsilon v}$
and in (1d) for the effective couplings of the type $g_{Fjs}$.
The dashed lines stand for any of the vector/axial mesons
and the dot-dashed lines for any of the tensors 
${\cal F}_{\mu\nu}, {\cal F}_{\mu\nu}^i, {\cal G}_{\mu\nu},
{\cal G}_{\mu\nu}^i$.
The different thickness of the dashed and dot-dashed 
 lines stand for the possibility of 
the couplings between different vector or axial mesons, 
i.e. in the case these diagrams can be seen as
vector/axial  mesons  mixings due to the
interaction with a constituent quark current.

\begin{figure}[ht!]
\centering
\includegraphics[width=100mm]{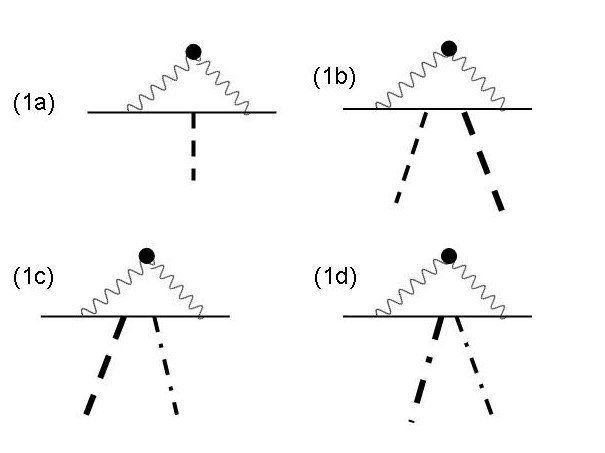}
\caption{ \label{fig:diagrams}
\small
In these diagrams, the wavy line
 with a full dot is a (dressed) non perturbative gluon propagator,
the solid  line represents the constituent quark,
the dashed line represents a vector or axial meson 
 $V_\mu, V_\mu^i, \bar{A}_\mu, \bar{A}_\mu^i$ and
dot-dashed line represents a strength tensor line 
for a vector/axial  meson
${\cal F}_{\mu\nu}, {\cal F}^i_{\mu\nu},{\cal F}^A_{\mu\nu},{\cal F}^{i,A}_{\mu\nu}$.
Diagram (1a) stands for the couplings $g_{r1}$ and $g_{v1}$.
Diagrams (1b), (1c) and (1d) represent respectively 
effective couplings with coupling constants $g_{va-j}$, $g_{\epsilon v}$
and $g_{Fjs}$
from expression (\ref{2v-j}).
In diagrams (1b,1c,1d) the vector  or axial  mesons lines might be the same 
or 
they might  be different  in the case of  mixings,
therefore the corresponding lines were drawn with different thickness.
}
\end{figure}

\subsection{Free vector and axial  mesons terms}

Although the aim of this work is to investigate
vector/axial mesons interactions with constituent quarks it is interesting
to have in mind some leading terms emerging from 
the effective action (\ref{Idet}) for the strict vector/axial mesons
sector.
This sector
has been investigated 
in some works within very similar formalisms for the NJL model 
\cite{meissner,ER-1986} 
and expression (\ref{gf-vecmes})
 below 
contains  basically the same vector mesons free terms.
The  following leading free vector/axial  mesons terms arise
in the very longwavelength limit
for zero momentum exchange:
\begin{eqnarray} \label{Ifree}
I_{free} &=&
- \frac{g_f^{(0)}}{4} \left( {\cal F}^{\mu\nu}_i {\cal F}_{\mu\nu}^i 
+ 
{\cal G}^{\mu\nu}_i {\cal G}_{\mu\nu}^i
+ 
{\cal F}^{\mu\nu} {\cal F}_{\mu\nu}
+
{\cal G}^{\mu\nu} {\cal G}_{\mu\nu}
\right)
-
\frac{M_v^{(0)}}{2} \left( {V_\mu^i}^2 + {\bar{A}_{i,\mu}}^2
+
{V_\mu}^2
+ {\bar{A}_\mu}^2
\right),
\end{eqnarray}
where the following effective parameters have been defined:
\begin{eqnarray} \label{gf-vecmes}
g_{f}^{(0)} 
&=&
  i d_1 4 N_c \; Tr' \; 
(( \tilde{S}_0 (k) \tilde{S}_0 (k))),
\\  \label{mass-vecmes}
{M_v^{(0)}}^2
 &=& 
- i d_1 8  N_c \; Tr' \; 
(( {S}_0 (k) S_0 (k) )).
\end{eqnarray}
It is  interesting to note that 
both of the two quantities, $g_f$ and $M_v^2$,
are the same for all the vector and axial mesons.
They
remain non zero
in the  chiral limit 
of zero quark effective mass $M^* \to 0$.
In this formulation the vector/axial mesons masses 
are all the same for the four mesons (\ref{mass-vecmes})
and the normalization constants of the canonical meson field definitions are also 
the same.
 Therefore these
expressions only satisfy one of the Weinberg sum rules,
$f_V^2 m_V^4 = f_A^2 m_A^4$,
 due to the absence of the coupling to pions 
 \cite{andrionov-espriou}. 
The complete resulting vector and axial meson sector with 
leading self interactions 
will be presented elsewhere.

\section{Numerical results}

The expressions for the effective coupling constants depend on
components of the gluon and quark propagator.
However it is possible to find
exact and  approximated  ratios 
between them that  provide approximated  estimations of 
their relative strengths.
The limit of very large quark effective mass might  be
obtained, for example,  by observing  that 
$S_0 (k) \sim 1/M^*$ and $\tilde{S}_0 (k)  \sim 1/{M^*}^2$.
 This yields the following approximated ratios:
\begin{eqnarray} \label{ratio-1}
\frac{g_{va-j}}{g_{r1}} \sim  \frac{3}{ 2 M^* } ,  && 
\;\;\;\;\;\;\;\;
\frac{g_{Fjs}}{g_{r1}} \sim \frac{3}{ 2 {M^*}^3}
  .
\end{eqnarray}
These ratios show that the   gauge-type  single 
vector mesons    couplings to  quark currents $g_{r1}$
are   the leading ones as compared to the others 
 in   the large quark mass expansion presented above
as it should be.
There is also an exact  ratio between coupling constants 
that  is given by:
\begin{eqnarray} \label{ratio-2}
\frac{g_{\epsilon v}}{g_{va-j}} = - \frac{1}{ M^* }
.
\end{eqnarray}
This exact ratio has the shape of a gauge invariant relation
since it does not depend on the gluon propagator,
however  it has been found by considering  (\ref{gluonprop}).

In Table 1 numerical values  are presented for  the effective coupling constants 
of expressions (\ref{gr1},\ref{gva1},\ref{Fjs})
and also for the parameters (\ref{gf-vecmes},\ref{mass-vecmes})
for 
 different  values of  the quark effective mass.
Two gluon propagators were chosen, 
one of them with only a 
transversal component from Tandy-Maris $D_{I}(k)$
\cite{SD-rainbow,gluon-pro}
and  the other is an  effective longitudinal  one  by Cornwall 
$D_{II}(k)$
 \cite{cornwall}.
Both of them are written below, they
yield dynamical chiral symmetry breaking
and the following association was adopted:
\begin{eqnarray} \label{propagators}
g^2 \tilde{R}^{\mu\nu} (k) \equiv  h_a D_i^{\mu\nu} (k)
\end{eqnarray}
where $D^{\mu\nu}_i(k)$ ($i=I,II$) 
 is one of the chosen  gluon propagators
from the  quoted articles, $h_a$ is a constant factor
which corresponds to fix the quark gluon (running) 
coupling constant.
The value of this factor $h_a$ was chosen to make the
coupling constant $g_{v1}=g_{r1}$ to reproduce a typical 
numerical value considered in nucleon nucleon potential \cite{rho-omega}
 in the vacuum 
or from nuclear properties. 
For example  from the  KSRF relation \cite{rho-omega,SWbook,djukanovic}
in the vacuum can be written as $g^2 = \frac{M_\rho^2}{2 F_\pi^2}$, where
$M_\rho = 770$ MeV and $F_\pi = 92$ MeV, it yields $g \simeq 6$.
From quark meson coupling model in different approximations
these in medium coupling constants have   values 
in a range $4.2 \leq g_\rho \leq 8.5$ and $6.8 \leq g_\omega \leq 9.5$
 \cite{qmc-coupling}. 
The fixed chosen value was  $g_{v1} h_a = 12$.
The overall normalization and momentum dependence 
of the gluon  propagators are different 
and therefore they provide considerably different values
for the resulting vector mesons-constituent quark coupling constants.
The expressions for the gluon propagators considered 
below are the following:
\begin{eqnarray}
D_I (k) &=& 
\frac{8  \pi^2}{\omega^4} De^{-k^2/\omega^2}
+ \frac{8 \pi^2 \gamma_m E(k^2)}{ \ln
 \left[ \tau + ( 1 + k^2/\Lambda^2_{QCD} )^2 
\right]}
,
\\
D_{II} (k) &=& 
K_F/(k^2+ M_k^2)^2 ,
\end{eqnarray} 
where
for the first expression 
$\gamma_m=12/(33-2N_f)$, $N_f=4$, $\Lambda_{QCD}=0.234$GeV,
$\tau=e^2-1$, $E(k^2)=[ 1- exp(-k^2/[4m_t^2])/k^2$, $m_t=0.5 GeV$,
$\omega = 0.5$GeV, $D= 0.55^3/\omega$ (GeV$^2$) \cite{SD-rainbow,gluon-pro}; 
and for the second expression
$K_F = (  2 \pi  M_k / (3 k_e) )^2$
where  $k_e  = 0.15$ was chosen together with 
the value of $h_a$ and $M_k = 220$MeV \cite{cornwall}.
The numerical results for the free vector mesons parameters $g_f^{(0)}$ and 
$M_v^{(0)}$  are 
ultraviolet (UV) divergent and therefore a 
momentum cutoff was considered.
However these two parameters 
 cannot be expected to reproduce experimental data  due to the 
limit of structureless mesons considered in this work.
When comparing the numerical  values exhibitted in the Table
it is seen they do not satisfy the ratios estimated above (\ref{ratio-1},\ref{ratio-2})
because the effective quark mass $M^* \sim 330$MeV is not 
large enough to reproduce the analytical ratios above. 
Larger values of the effective mass 
 however are not realistic and they were not included 
in the Table.
Nevertheless 
it can be noted that the resulting numerical values for 
 larger values quark effective mass in the Table are closer to the
approximated ratios estimated above.

\begin{table}[ht]
\caption{
\small  In the first column the quark effective masses are displayed with
the values of the factor $h_a$ that were chosen to fix  the
coupling constant $g_{v1}=g_{r1} = 12$
 as a  typical 
numerical value considered in nucleon nucleon potential \cite{rho-omega}.
 In the second column the gluon propagators are  indicated:
$D_I(k)$ and $D_{II}(k)$ are  the gluon propagators respectively 
from Refs. \cite{SD-rainbow,gluon-pro} and  
Ref. \cite{cornwall}.
In the other columns,
 results,   that depend on the gluon propagator, 
 from the expressions (\ref{gr1},\ref{gva1},\ref{Fjs})
are displayed
and also  parameters  do not depend on the gluon propagator 
(\ref{gf-vecmes}) and (\ref{mass-vecmes}).
The momentum cutoff used for the integrations (\ref{gf-vecmes},\ref{mass-vecmes}) 
 is indicated together with $g_f^{(0)}$.
} 
\centering % used for centering table
\begin{tabular}{c c c c c c c c } % centered columns (columns)
\hline\hline %inserts double horizontal lines
$M^*$ , $h_a$ & $D_i (k)$ &  
 $g_{r1}   h_a$    &  $g_{vaj}  h_a$   
%& $g_{FV} h_a$  
  & $g_{Fjs} h_a$
%% $g_{qA} $  & $g_1$ &  $g_2$ & $g_{3}$ 
& 
$g_{f}^{(0)}$ ($\Lambda$) 
& 
$M_v^{(0)}$ 
\\
 (GeV), -  &  &         &  (GeV$^{-1}$) 
%  &  (GeV$^{-2}$) 
& (GeV$^{-3}$)  
& \; -   (GeV) \;
& (GeV)
\\
\hline
% \\ [0.5ex]
0.33, $\frac{12}{9.3}$  & $D_I$  & 12 &  5.8   & 111 & 0.10  (2.0) &  0.479 
\\ [0.5ex]
0.33,  $\frac{12}{0.67}$ & $D_{II}$  &  12  &  5.7   & 107 & - & - 
\\ [0.5ex]
0.28, $\frac{12}{10.5}$  & $D_I$  & 12 &   6.4  &  165  & 0.11 (2.0) & 0.495 
\\ [0.5ex]
0.28,  $\frac{12}{0.75}$ & $D_{II}$  &  12  & 6.4 &  163   &  - & -
\\ [0.5ex]
0.22,  $\frac{12}{12.7}$  & $D_I$  & 12  &  6.8   & 279 &  0.13 (2.0)  &  0.512
\\ [0.5ex]
0.22,  $\frac{12}{0.9}$ & $D_{II}$  & 12  &  6.7  &  285 & - & - 
\\ [0.5ex]
0.07,  $\frac{12}{20.3}$  & $D_I$  &  12  & 7.2  & 4944 &  0.22 (2.0)&  0.545
 \\ [0.5ex]
0.07,  $\frac{12}{1.5}$ & $D_{II}$  & 12    &  7.6  &  3228 & - & -
\\[1ex] % [1ex] adds vertical space
\hline %inserts single line
\end{tabular}
\label{table:results-2} % is used to refer this table in the text
\end{table}

\subsection{ Form factors}

The complete expressions for two of the form factors
$g_{r1}$ and $g_{va-j}$,  expressions
(\ref{gr1}) and (\ref{gva1}), 
will be  generalized  for non zero momentum transfer.
To explain the notation,   two examples of 
 full momentum dependent terms 
in expressions (\ref{vec-quark}) and (\ref{2v-j}) 
are shown corresponding to a particular channel of the diagrams
shown in Fig. \ref{fig:diagrams}. 
 They can be written after a Fourier transformation  as:
\begin{eqnarray} 
{\cal L}_{ff} = 
g_{r1} (Q) V_\mu^i (Q) \bpsi (Q) \gamma^\mu \sigma^i \psi (0)
+ g_{va-j} (Q_1,Q_2) 
V_\mu^i (Q_1) V^\mu (Q_2)  \bpsi (Q_1+Q_2)  \sigma^i \psi (0)
.
\end{eqnarray}
The couplings $g_{va-j}$ in expression
 (\ref{2v-j}) are the same  for two identical vector/axial mesons couplings
to quarks and for two different vector/axial mesons couplings to quarks
although the quark currents might be  different in each case.
After a Wick rotation to the Euclidean momentum space, 
the form factors  were calculated numerically.
For the two-vector mesons couplings to constituent quarks two different
calculations were performed, a complete one ($^{com}$) and a momentum-truncated one
($^{tr}$).
The truncated expression is  obtained by the following 
approximation: $S_0(k) \simeq M^* \tilde{S}_0(k)$.
The following  expressions  
were investigated numerically:
\begin{eqnarray} 
  \label{ff-gr1-tr}
 g_{r1} (Q) &=& 
 \;  4 N_c \; d_1 (\alpha g^2) {M^*}^2 \;
  \int_k 
\tilde{S}_0 (k)
\tilde{S}_0 (k+ Q) 
 \bar{\bar{R}} (-k ) 
,
\\  \label{ff-gvaj-co}
g_{va-j}^{com} (Q_1, Q_2) &=&
12 \; N_c \; d_2  (\alpha g^2)  {M^*}  \;
 \int_k
T_{Q_1,Q_2}(k)
\tilde{S}_0 (k) \tilde{S}_0 (k+ Q_1)
\tilde{S}_0 (k+ Q_1+ Q_2) 
{{R}}(-k )
,
\\
\label{ff-gvaj-tr}
g_{va-j}^{tr} (Q_1, Q_2) &=&
12 \; N_c \; d_2  (\alpha g^2)   {M^*}^3 \;
 \int_k
\tilde{S}_0 (k) \tilde{S}_0 (k+ Q_1)
\tilde{S}_0 (k+ Q_1+ Q_2) 
{R}(-k )
,
\end{eqnarray}
where 
 $\int_k = \int d^4 k/(2\pi)^4$, 
and the momentum dependent 
functions above 
 are given by:
\begin{eqnarray}
\tilde{S}_0 (k) &=& \frac{1}{k^2 + {M^*}^2}  ,
\\
\tilde{S}_2 (k,k+Q) &=&
\frac{ k^2 + k \cdot Q - {M}^2 
}{ (k^2 + {M^*}^2) ( (k+Q)^2 + {M^*}^2)} ,
\\
T_{Q_1,Q_2} (k) &=& \left[
3 k^2 + 4 k\cdot Q_1 + 2 k \cdot Q_2 + 
Q_1 \cdot Q_2 + Q_1^2 - {M^*}^2
\right],
\end{eqnarray}
and $\bar{\bar{R}}(k)$ was given after expression (\ref{gr1}).

\begin{figure}[ht!]
\centering
\includegraphics[width=120mm]{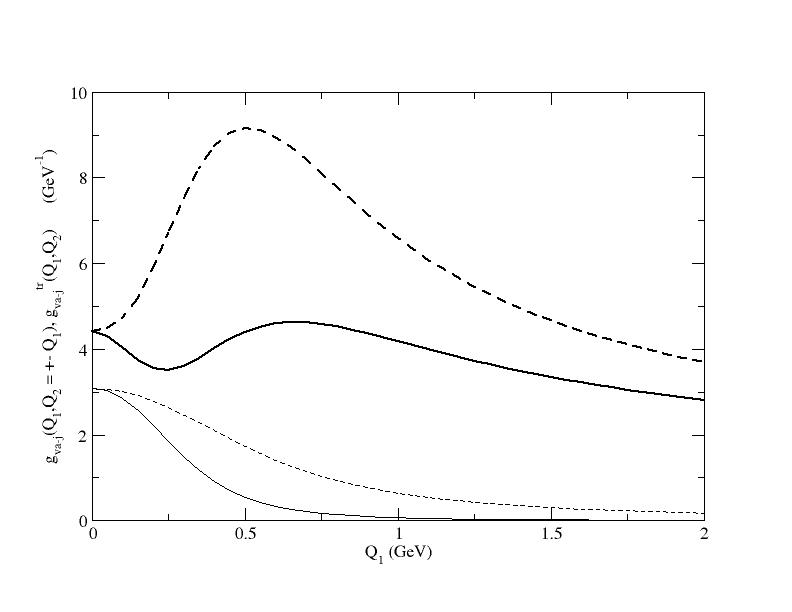}
\caption{ \label{fig:ffvaja-qq-TM}
\small
% ffvaja-I-qq-s2+tr-TM.jpg
By considering the gluon propagator I from Ref. \cite{SD-rainbow,gluon-pro}
the expressions for  $g_{va-j}^{com} (Q_1,Q_2=+ Q_1)$
and 
$g_{va-j}^{com} (Q_1,Q_2=- Q_1)$
 are plotted respectively  in solid thick line and dashed thick line;
$g_{va-j}^{tr} (Q_1,Q_2= + Q_1)$
and 
$g_{va-j}^{tr} (Q_1,Q_2=- Q_1)$
 are plotted respectively  in solid  line and dashed  line.
}
\end{figure}

\begin{figure}[ht!]
\centering
\includegraphics[width=120mm]{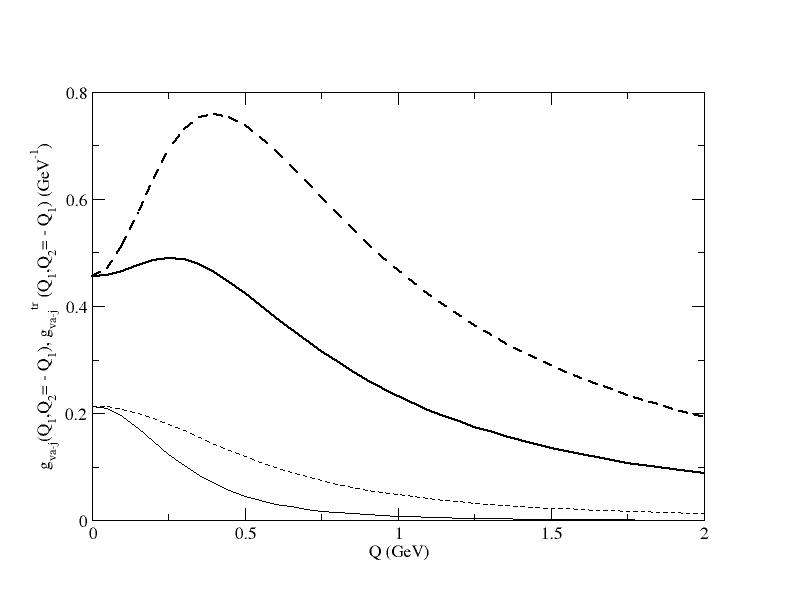}
\caption{ \label{fig:ffvaja-qq-CO}
\small
% ffvaja-I-qq-s2+tr-CO.jpg
By considering the gluon propagator II from Ref. \cite{cornwall}
the expressions for  $g_{va-j}^{com} (Q_1,Q_2=+ Q_1)$
and 
$g_{va-j}^{com} (Q_1,Q_2=- Q_1)$
 are plotted respectively  in solid thick line and dashed thick line;
$g_{va-j}^{tr} (Q_1,Q_2= + Q_1)$
and 
$g_{va-j}^{tr} (Q_1,Q_2=- Q_1)$
 are plotted respectively  in solid  line and dashed  line.
}
\end{figure}

\begin{figure}[ht!]
\centering
\includegraphics[width=120mm]{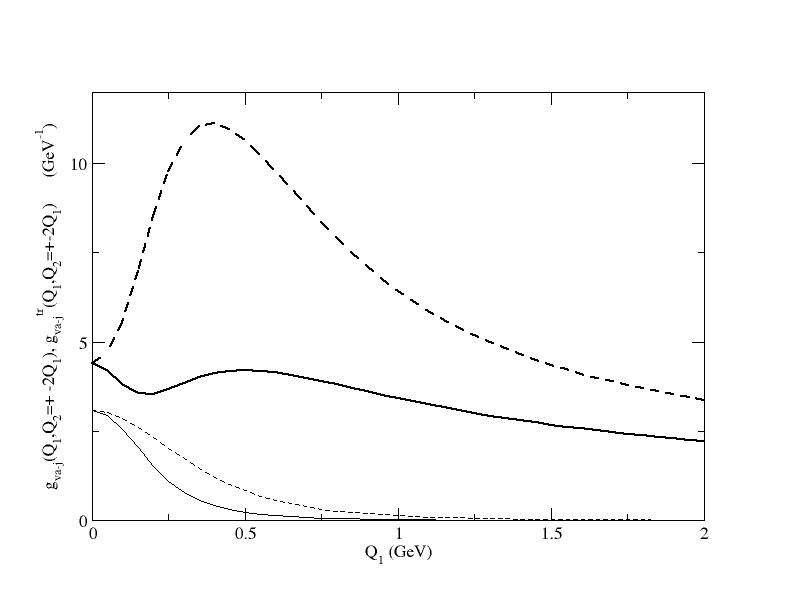}
\caption{ \label{fig:ffvaja-I-q2q-TM}
\small
% ffvaja-I-q2q-s2+tr-TM.jpg
By considering the gluon propagator I from Ref. \cite{SD-rainbow,gluon-pro}
the expressions for  $g_{va-j}^{com} (Q_1,Q_2=+ 2 Q_1)$
and 
$g_{va-j}^{com} (Q_1,Q_2=- 2 Q_1)$
 are plotted respectively  in solid thick line and dashed thick line;
$g_{va-j}^{tr} (Q_1,Q_2= +2  Q_1)$
and 
$g_{va-j}^{tr} (Q_1,Q_2=- 2 Q_1)$
 are plotted respectively  in solid  line and dashed  line.
}
\end{figure}

\begin{figure}[ht!]
\centering
\includegraphics[width=120mm]{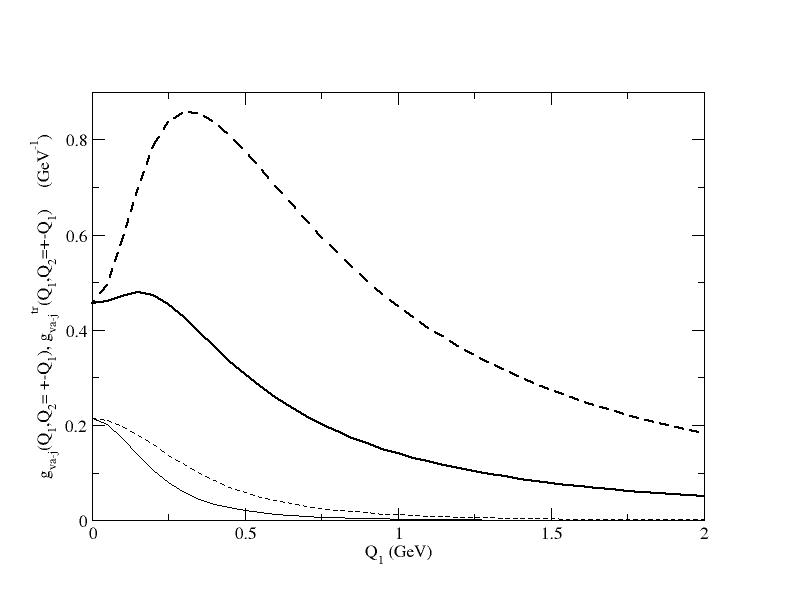}
\caption{ \label{fig:ffvaja-I-q2q-CO}
\small
% ffvaja-I-q2q-s2+tr-CO.jpg
By considering the gluon propagator II  from Ref. \cite{cornwall}
the expressions for  $g_{va-j}^{com} (Q_1,Q_2=+ 2 Q_1)$
and 
$g_{va-j}^{com} (Q_1,Q_2=- 2 Q_1)$
 are plotted respectively  in solid thick line and dashed thick line;
$g_{va-j}^{tr} (Q_1,Q_2= +2  Q_1)$
and 
$g_{va-j}^{tr} (Q_1,Q_2=- 2 Q_1)$
 are plotted respectively  in solid  line and dashed  line.
}
\end{figure}

\begin{figure}[ht!]
\centering
\includegraphics[width=120mm]{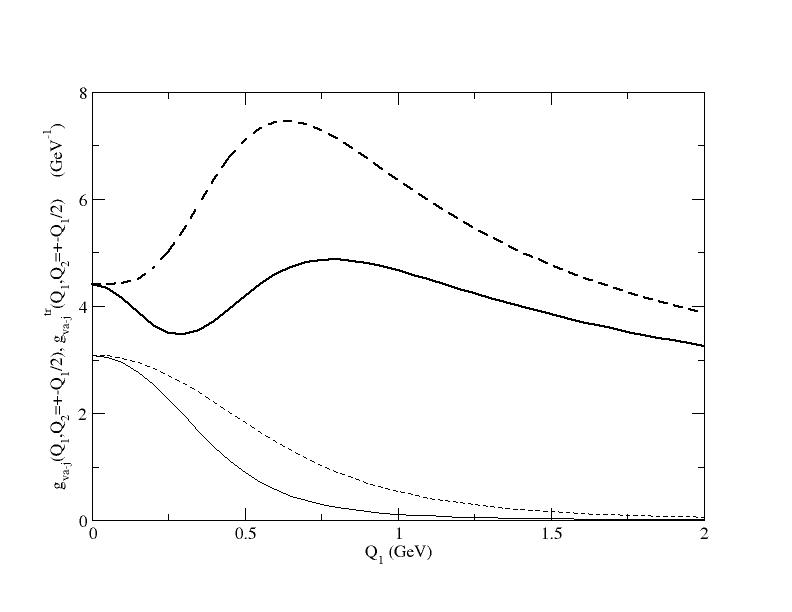}
\caption{ \label{fig:ffvaja-I-qq2-TM}
\small
% {ffvaja-I-qq2-s2+tr-TM.jpg}
By considering the gluon propagator I from Ref. \cite{SD-rainbow,gluon-pro}
the expressions for  
$g_{va-j}^{com} (Q_1,Q_2=+  Q_1/2)$
and 
$g_{va-j}^{com} (Q_1,Q_2=-  Q_1/2)$
 are plotted respectively  in solid thick line and dashed thick line;
$g_{va-j}^{tr} (Q_1,Q_2= + Q_1/2)$
and 
$g_{va-j}^{tr} (Q_1,Q_2=-  Q_1/2)$
 are plotted respectively  in solid  line and dashed  line.
}
\end{figure}

\begin{figure}[ht!]
\centering
\includegraphics[width=120mm]{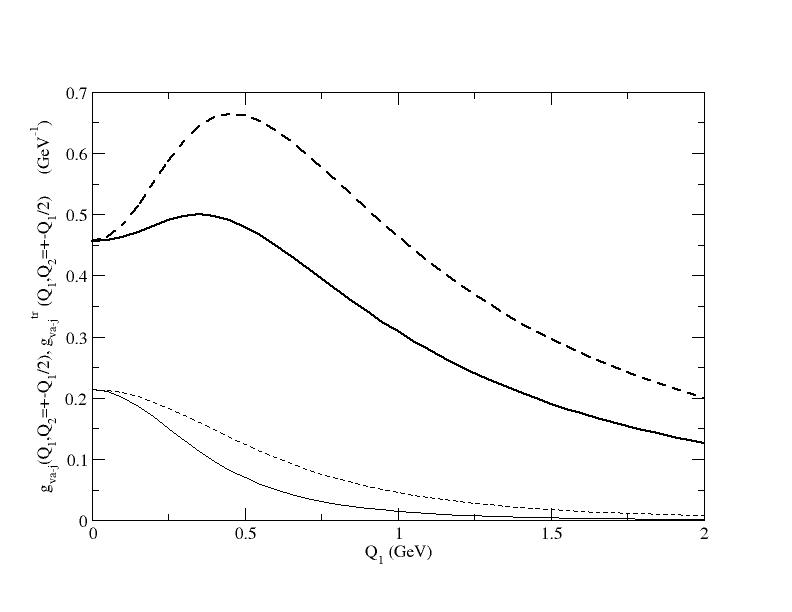}
\caption{ \label{fig:ffvaja-I-qq2-CO}
\small
% {ffvaja-I-qq2-s2+tr-CO.jpg}
By considering the gluon propagator II from Ref. \cite{cornwall}
the expressions for 
$g_{va-j}^{com} (Q_1,Q_2=+  Q_1/2)$
and 
$g_{va-j}^{com} (Q_1,Q_2=-  Q_1/2)$
 are plotted respectively  in solid thick line and dashed thick line;
$g_{va-j}^{tr} (Q_1,Q_2= + Q_1/2)$
and 
$g_{va-j}^{tr} (Q_1,Q_2=-  Q_1/2)$
 are plotted respectively  in solid  line and dashed  line.
}
\end{figure}

In Figures 2-7 the form factors 
$g_{va-j}^{com}(Q_1,Q_2),g_{va-j}^{tr}(Q_1,Q_2)$ 
of  expressions  above
are presented as functions of $Q_1$
for the two   gluon propagators $D_I,D_{II}$ and 
different external momenta $Q_2$ by considering  $M^*=330$MeV.
The cases for two incoming vector/axial  mesons to the vertex
in the same direction, $Q_1 \cdot Q_2 = |Q_1| |Q_2| > 0$, are presented
with solid lines and  
the case for one incoming  meson and  another outgoing meson ($Q_2 < 0$
and $Q_1 \cdot Q_2 = |Q_1| |Q_2| <  0$)
from the interaction vertex,
in the same longitudinal direction,
 are presented with dashed lines in all these Figures.
The thick lines correspond to the complete expressions ($^{com}$) and the
thin lines to the truncated ones $^{tr}$.
Figures (\ref{fig:ffvaja-qq-TM},\ref{fig:ffvaja-I-q2q-TM},\ref{fig:ffvaja-I-qq2-TM}) 
are drawn with propagator $D_I(k)$
and Figures (\ref{fig:ffvaja-qq-CO},\ref{fig:ffvaja-I-q2q-CO},\ref{fig:ffvaja-I-qq2-CO})
 are drawn with  $D_{II}(k)$.
In Figs. (\ref{fig:ffvaja-qq-TM}) and (\ref{fig:ffvaja-qq-CO}) 
 it was considered  
$Q_2 = \pm Q_1$, 
whereas in Figs. (\ref{fig:ffvaja-I-q2q-TM}) and (\ref{fig:ffvaja-I-q2q-CO})
$Q_2 =  \pm 2 Q_1$ and
finally in Figs. (\ref{fig:ffvaja-I-qq2-TM}) and (\ref{fig:ffvaja-I-qq2-CO})
 $Q_2 = \pm Q_1/2$.
In all these figures it is seen that the more intrincated momentum structures of 
 $g_{va-j}^{com}(Q_1,Q_2)$
 produce a non monotonic behavior
whereas the truncated expression yields a
much faster decreasing behavior of the form factor with
increasing $Q_1$.
In all the figures,  the numerical
values for mesons with  opposite momenta 
$ Q_2 < 0$ are always larger than the parallell incoming mesons
$ Q_2 > 0$.
 The former, in all cases,   
go to zero considerably slower than the latter.
For $Q_2 = + \frac{a}{2} Q_1$  
 ($a=1,2$ and $4$ respectively  in  Figures 
(\ref{fig:ffvaja-I-qq2-TM},\ref{fig:ffvaja-I-qq2-CO}), 
(\ref{fig:ffvaja-qq-TM},\ref{fig:ffvaja-qq-CO}) 
and (\ref{fig:ffvaja-I-q2q-TM},\ref{fig:ffvaja-I-q2q-CO}) )
the complete expressions exhibit quite different behaviors with  low momentum 
when comparing the results from propagators $I$ and $II$.
The complete expression  with propagator $I$ (Figs. 2,4,6) in thick solid lines
yields a local minimum for low momentum $Q_1$ that turns out to be always close 
to zero in the case of the results with propagator $II$.
It is also noted that the propagator $II$ yields results that go to zero faster
with $Q_1$ than the calculation with propagator $I$.
For the  behavior of $g_{va-j}^{com}(Q_1,Q_2)$
with negative  $Q_2 = - \frac{a}{2} Q_1$ ($a=1,2,4$ in the Figures), 
there is a maximum value of $g_{va-j}^{com}$ that occurs 
 in larger values of $Q_1$ for
smaller  $a$ (or smaller $Q_2$). 
For example, by comparing the thick dashed lines 
 in Figures 4 and 6, respectively
  for $Q_2=- 2 Q_1$  and  $Q_2 = - Q_1/2$,  
the maximum value  appears to be respectively 
around $Q_1 \simeq 400$MeV 
and $Q_1 \simeq 650$MeV for propagator $I$.



\begin{figure}[ht!]
\centering
\includegraphics[width=120mm]{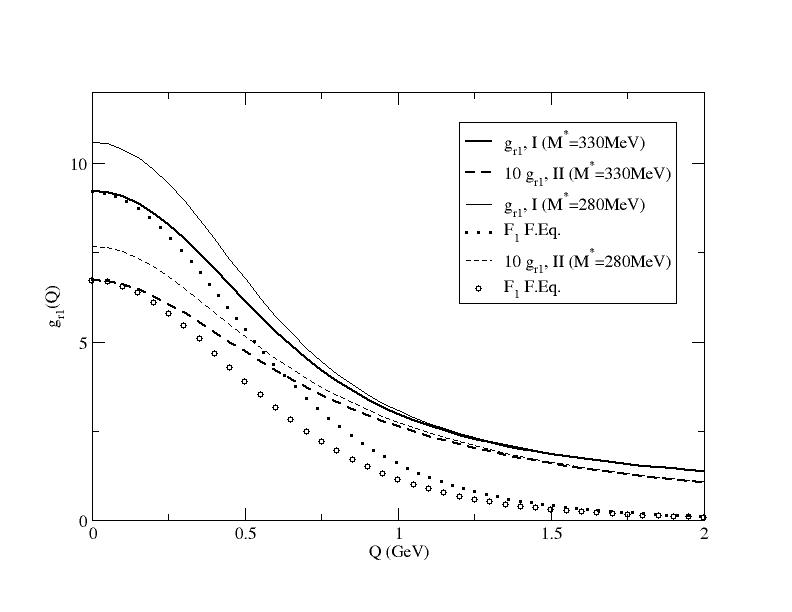}
\caption{ \label{fig:gr1-TM+CO}
\small
%gr1-330-280+CO+TM.jpg
The strong vector meson form factor $g_{r1}(Q)$
given in expression (\ref{ff-gr1-tr}) 
is presented for the two gluon propagators I \cite{SD-rainbow,gluon-pro} 
- solid lines - 
and II \cite{cornwall} - dashed lines.
The quark effective mass was considered $M^*=330$MeV for 
thick lines and $M^*=280$MeV for thin lines.
The normalized vector meson-nucleon  form factor from Ref. \cite{vec-nuc-craig-etal}
is also presented for further comparison
with corresponding normalization with results of the two propagators
 above:
$F_1^\rho (0)=  9.2$ (propagator I, thick dotted line)
and  $F_1^\rho (0) = 6.7$ (propagator II, circles).
}
\end{figure}

In Figure 8  the form factor 
$g_{r1} (Q)$  is  plotted 
for  $M^*=330$MeV 
 with 
 gluon propagators $D_I$ and $D_{II}$
respectively  in 
thick solid and dashed lines. 
By 
considering  $M^*=280$MeV  
the same convention  was adopted 
for thin  solid and thin  dashed lines.
The results for $M^*=330$MeV
 are compared to the   nucleon vector form factor
fitted by a quadrupolar form Ref.
 \cite{vec-nuc-craig-etal}
from Faddeev equation
in dotted line and circles. 
This fit is given by expression:
$F_{1}^{\rho} (Q^2) = \frac{F_1^\rho (0)}{ 
\left(1 + \frac{Q^2}{\Lambda_{1,\rho}^2} \right)^3}$,
where  $\Lambda_{1,\rho}=1.12$ {GeV}, 
and
$F_1^\rho (0) = g_{\rho NN}$  is in the range $4.82 - 6.4$
in different works.
To allow the comparison of the strict momentum dependence  
the value of $F_1^V(0)$ was taken to be 
the numerical zero momentum form factors obtained in this work. 
Therefore   $F_1^\rho (0)=  9.2$ (propagator I, thick dotted line)
and  $F_1^\rho (0) = 6.7$ (propagator II, circles).
The large-$Q$ behavior of $g_{r1}(Q)$
 is dictated by the quark effective mass as it can be 
seen in the region of $Q=2$GeV.

The strong vector meson  form factors yields 
the  strong squared radius for the vector and axial mesons.
With the expression (\ref{ff-gr1-tr})
 the following expression  was calculated:
\begin{eqnarray} \label{rqm}
< r_\rho^2 >_s  &=& - \left. 6 
\frac{d g_{r1} (Q)}{d Q^2} 
 \right|_{Q=0}
.
\end{eqnarray}
An exact relation  between this strong squared radius (\ref{rqm})
and the electromagnetic squared radius for the 
rho and omega vector mesons, $<r_\rho^2>_E$ (in expression (41) of 
Reference \cite{FLB-2017c})
and  $<r_\omega^2>_E$ (both for zero magnetic field)
follows and it is given by:
\begin{eqnarray} \label{rqm-S-E}
<r_\rho^2 >_s 
= 
\frac{2}{ e} <r_\rho^2>_E  ,
\;\;\;\;\;\;\;\;\;
<r_\omega^2 >_s 
= 
\frac{2}{3 e} <r_\omega^2>_E. 
\end{eqnarray}
The rho quadratic radius (\ref{rqm})
is presented in Fig. 9  for the two gluon propagators 
respectively $D_I$ and $D_{II}$.
Most of the numerical results are of the order of magnitude
of the experimental data, $<r^2_\rho>_{exp} \simeq 0.28-0.56$fm$^2$
\cite{exp-rhoqm,holo-krein-etal},
except the numerical results obtained with  the gluon propagator $D_I$
for which  the values are divided  by a factor 10 to keep the scale of the figure.
These resuts manifest the strong dependence of the results 
on the strength of the quark-gluon coupling and on the overall momentum
dependence of the 
gluon propagator.

\begin{figure}[ht!]
\centering
\includegraphics[width=120mm]{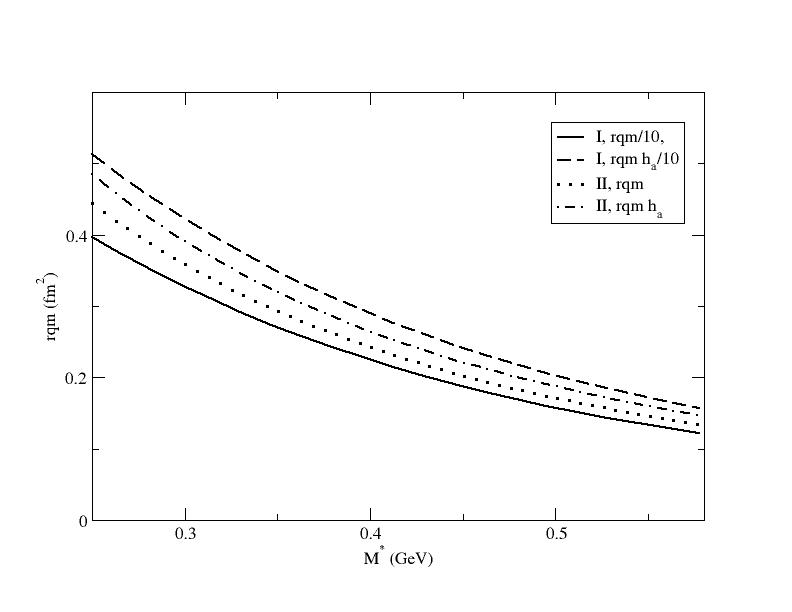}
\caption{ \label{fig:rqm-I}
\small
By considering propagators I and II 
the 
vector meson strong quadratic radius  
is exhibitted as a function of the quark effective mass $M^*$.
Results are shown 
with and without the factor $h_a$ explained in the Table:
with  thick solid and thick dashed lines with $D_I(k)$ 
and they are multiplied by $1/10$ to be kept in the scale of the Figure.
The cases for gluon propagator propagator $D_{II}$
are represented in
 dotted and dot-dashed lines respectively 
with and without $h_a$.
}
\end{figure}

 \section{ Summary and final remarks}

To summarize,  different effective couplings 
between
 vector/axial mesons and constituent 
quarks were presented.
The effective coupling constants were obtained as the zero momentum limit
of form  factors  and they were
expressed in terms of the parameters of the original  model  and 
of components of the quark and gluon kernels.
The vector and axial mesons fields were introduced by means of the 
auxiliary field method
and 
the $\omega$  and $f_1$  vector and axial mesons   were 
considered to be  chiral partners.
The structureless mesons limit was considered for the zero order derivative 
and large quark effective mass expansion.
The leading vector mesons-constituent quark
interactions are the minimal gauge couplings  in expression 
(\ref{vec-quark}).  
The determinant 
expansion  
given by expression (\ref{S0v})
 can be written in terms of a covariant derivative 
for the minimal coupling with vector/axial mesons.
However most of the 
two vector/axial mesons effective  couplings 
to constituent quarks 
(\ref{2v-j})
cannot be written in terms of non Abelian parts
of vector mesons strength tensors
 that  generalize  expressions (\ref{tensors})
and therefore 
 do not seem to be possibly  incorporated
into a chiral vector/axial mesons gauge framework 
\cite{meissner,birse,HLS}.
Some next leading, or second order, effective couplings 
can be associated to a  part of the 
elastic and  inelastic scattering amplitude of 
vector mesons-constituent quarks
scattering amplitude and some of the inelastic ones 
might be seen as vector or axial  mesons mixings mediated 
or induced by constituent 
quark currents exhibited  in expression (\ref{2v-j}).
They present considerably  different structures
 from the usual
charge symmetry violation  mixing.
 Also they emerge  not only 
for the neutral $\rho^0, A_1$ but also for the charged $\rho, A_1$-mesons
coupled to quark currents
such as $\vec{\rho}^\mu (\partial_\mu \omega_\nu) \; 
\cdot
\bpsi \vec{\sigma} \gamma^\nu \psi$. 
It is interesting to note that only one effective coupling constant was 
found to emerge in the leading order in expressions (\ref{vec-quark}),
and only three different coupling constants at the second order 
in expression (\ref{2v-j}) in spite of the relatively large number of
different effective couplings. 
This issue goes along the Universality idea.
However these effective couplings do  not represent
 all the possible effective couplings since
 tensor quark currents 
are not obtained from the  method presented in this work.
The resulting leading single meson couplings (\ref{vec-quark}) are
the only renormalizable effective couplings 
and the higher order 
ones, such as (\ref{2v-j}), are non renormalizable.
All the coupling constants, however,
 are UV finite for usual large momentum
behaviors of the 
gluon propagator such as $D(p) \sim 1/p^s$ for $s > 2$.
Furthermore each additional vector/axial meson
that appear in higher orders terms of the large quark mass expansion
will present progressively additional factors $S_0 (k) \sim 1/M^*$
and these extra factors will make the 
higher order coefficients
of the terms  (effective coupling constants) 
 of the expansion to be progressively smaller
in the large quark mass regime.
Numerical estimates were presented
by considering two very different 
gluon propagators. 
Although the resulting  order of magnitude of the 
leading vector mesons-constituent quark coupling constants, $g_{v1}, g_{r1}$,
 nearly reproduced experimental or expected values,
these coupling constant $g_{v1}, g_{r1}$ were normalized to a 
typical value considered in the literature by fixing a particular value for
the  quark-gluon coupling constant. 
This was done by fixing $h_a$ as shown in the Table.
The other coupling constants of the Table,
for which we found no values in the literature,
 were corrected accordingly.
Eventually, it might be that, by considering  all the complementary 
mechanism(s)
from QCD for the couplings shown above - if there are relevant corrections
in different  QCD mechanisms -
the gauge independence should  be expected to be 
recovered  at the  hadron and nuclear levels.
Nevertheless, although the effective coupling constants shown above
have different dimensions, it was possible to estimate
approximated and exact  ratios between  them 
in the limit of large quark effective mass.
The momentum dependences of two different form factors, 
$g_{r1}(Q)$ and $g_{vqa-j}(Q_1,Q_2)$, were
 addressed  also
by  considering   a momentum truncation 
for the coupling $g_{va-j}$ in expression (\ref{ff-gvaj-tr}).
The momentum of the second vector/axial meson 
was chosen to assume the following values $Q_2 = \pm \frac{a}{2} Q_1$
for $a= 1, 2, 4$, i.e. its modulus  to be 
smaller, equal or larger than $Q_1$.
%{\bf The next interesting
%step is to calculate the resulting vector/axial meson-nucleon couplings and form factors
%from the constituent quarks case addressed in the present work
%to compare with experimental data.}
Finally the quark effective mass dependence of 
the strong  rho (or omega) square radius was investigated for the two gluon propagators.
 Pions dynamics and effective  couplings to constituent quarks were 
  presented in Ref. \cite{EPJA-2016}
and they make
 possible to consider constituent quark and vector mesons effective interactions
mediated by them.
They would correspond to a class of effective hadron interactions
mediated by pseudoscalar  auxiliary fields
that can be found by integrating out them approximatedly. 
With this, the resulting effective couplings would contain additional 
factors $S_0(k)$ or $\tilde{S}_0(k)$ being thereore  
of higher order in $1/M^*$  and therfore numerically  smaller.

\section*{Acknowledgment}

The author acknowledges  short discussions
 with  P. Bedaque, G.I. Krein 
and C.D. Roberts.

%\section*{References}

\end{document}